\documentclass[12pt]{article}

\usepackage{latexsym}
\usepackage{epsfig}
\newcommand{\sect}[1]{\setcounter{equation}{0}\section{#1}}

\def\be{\begin{equation}}
\def\ee{\end{equation}}
\def\ba{\begin{eqnarray}\samepage}
\def\ea{\end{eqnarray}}
\font\msbm= msbm10
\def\edth{\hbox{\msbm \char '147}}

\begin{document}

\title{{\bf Higher Spin Field Equations in a Virtual Black Hole Metric}}
 
\author{\\Tim Prestidge \\ \\{\sl T.Prestidge@damtp.cam.ac.uk} \\ \\ 
\\Department of Applied Mathematics and Theoretical Physics,\\The
University of Cambridge,\\Silver Street,\\Cambridge,\\CB3 9EW. \\ \\
{\bf damtp R-98/07}}

\maketitle

\begin{abstract}
\noindent
In a quantum theory of gravity, fluctuations about the vacuum may be
considered as Planck scale virtual black holes appearing and
annihilating in pairs. Incident fields scattering from such
fluctuations would lose quantum coherence.
 
In a recent paper Hawking \& Ross obtained an estimate for the
magnitude of this loss in the case of a scalar field. Their calculation
exploited the separability of the conformally invariant scalar wave
equation in the electrovac C metric background, which is justified as a
sufficiently good description of a virtual black hole pair in the
limit considered. 

In anticipation of extending this result, the Teukolsky equations for
incident fields of spin $s = 0$, $\frac{1}{2}$, $1$, and $2$ are
separated on the vacuum C metric background, and solved in the same
limit. These equations are shown in addition to be valid on the
electrovac C metric background when restricted to $s=0$, $\frac{1}{2}$,
and $1$. The angular solutions are found to reduce to the
spin-weighted spherical harmonics, and the radial solutions are found
to approach hypergeometrics close to the horizons. 

By defining appropriate scattering boundary conditions, these
solutions are then used to estimate the transmission and 
reflection coefficients for an incident field of spin $s$. The
transmission coefficient is required in order to estimate the loss of
quantum coherence of an incident field through scattering off virtual
black holes.

\end{abstract}

\renewcommand{\thepage}{ }
\pagebreak

\renewcommand{\thepage}{\arabic{page}}
\setcounter{page}{1}

\sect{Introduction}
The action for the gravitational field is {\sl not} scale invariant,
and so large fluctuations in the metric over short length scales will
not have a large action. Such fluctuations are therefore not precluded
since they are not damped in the path integral, and indeed even
variations in topology may occur since these would further increase
the associated action by only an arbitrarily small amount. Such
considerations led Wheeler~\cite{Wheeler} to suggest that in a quantum theory
of gravity, spacetime should be expected to have a `foam-like'
structure on scales of the Planck length or shorter, in which both the
metric and the topology of the manifold may vary considerably. On
larger scales however, spacetime would appear smooth and continuous,
reverting to the manifold structure more conventionally associated
with the treatment of classical general relativity.

Hawking~\cite{SWH1} and Hawking, Page \& Pope~\cite{HPP1,HPP2}
first interpreted this foam structure as made from `quantum
gravitational bubbles' arising in three topological varieties: $S^2
\times S^2$, $CP^2$, and $K3$. These would be the `building blocks' of
spacetime, -- not themselves solutions to any field equations but
rather occuring as quantum fluctuations. Their non-trivial topologies
do not allow complete foliation of the manifolds with a family of
non-intersecting time surfaces, and consequently Green functions
defined on them are bound to exhibit certain acausal and frequency
mixing properties more commonly associated with those defined on the
Euclidean black hole metrics. The interpretation of these bubbles as
virtual (or off-shell) black holes, which materialise out of the vacuum 
and subsequently vanish again, would therefore appear well justified. 

Quantum field theories defined on manifolds with non-trivial topology
are necessarily non-unitary, the consequences of which include the
loss of information and of quantum coherence. If spacetime is indeed
composed of Planck scale building blocks with non-trivial topologies,
then quantum fields propagating in it ought to suffer from these effects.
Such a notion may na\"{\i}vely appear to be somewhat at odds with
observation! Theories relying on unitary evolution and conserved
quantum coherence have been demonstrated to high degrees of accuracy,
and yet the quantum bubbles picture would seem to advocate the
loss of quantum coherence for fields propagating in spacetime due to
their unavoidable interaction with Planck scale fluctuations. How may
these apparent contradictions be reconciled? 

Calculations in simple $S^2 \times S^2$ and $CP^2$ models~\cite{HPP2} 
indicated that the non-trivial topologies of the quantum bubbles
induced additional singularities into the Green functions for particle
propagators. These tended to make large contributions to the {\it S}-matrix
in the scalar case, but only very small contributions in higher spin
cases for energies low with respect to the Planck scale. In a sense
then, the reconciliation is complete since all observed elementary 
particles have spin greater than zero. It is therefore hardly surprising 
that spacetime should appear smooth and nearly flat at current
observational energies, since the only available probes of its
structure are essentially blind to its short scale fluctuations.   

Consideration of the pair creation of black holes led
Hawking~\cite{SWH2} to reconsider the original interpretation of
spacetime foam. Gibbons~\cite{GWG1} had suggested that the pair creation 
of real black holes in a background field could be described by the
Ernst solution~\cite{ERNST} and its dilaton generalisations,-- the
Euclidean section of which has topology $S^2 \times S^2 - \{\mathrm
{point}\}$. By analogy with the pair creation of ordinary particles,
this topology corresponds to a black hole loop in a spacetime
asymptotic to $R^4$. However, $S^2 \times S^2 - \{\mathrm
{point}\}$ may also be considered as the topological sum of the compact
space $S^2 \times S^2$ and the non-compact space $R^4$, leading to a
reinterpretation of the $S^2 \times S^2$ bubbles in the spacetime foam
as closed loops of virtual black holes. Thus, rather than a single
off-shell black hole, the quantum bubbles are to be thought of as
virtual black holes appearing and annihilating in pairs. If this
should occur, then it seems plausible that incident particles could be
absorbed and subsequently re-emitted -- possibly as different
particles, and with loss of quantum coherence.

Adopting the perspective that the compact $S^2 \times S^2$ bubbles
occur as quantum fluctuations unaffected by incident low energy
particles, a consistent scheme for estimating the scattering amplitude
and loss of coherence may be constructed~\cite{SWH2}. This involves
computing the Euclidean Green function in an asymptotically Euclidean
metric on $S^2 \times S^2 - \{\mathrm{point}\}$, analytically
continuing to the Lorentzian section at infinity, and integrating with
the initial and final state wave functions of the scattered
field. Weighting the result according to the action of the Euclidean
metric gives the scattering amplitude, which should then be integrated
over all asymptotically Euclidean metrics. However this scheme is 
effectively intractable since neither the calculation of the Green
function in a general metric or the path integration over all such
metrics is possible.

This problem can be at least partially circumvented by demonstrating
the effects of coherence loss in any one set of appropriate metrics,
since other metrics within the path integral cannot then restore it. This
has been carried out for incident scalar fields by Hawking \&
Ross~\cite{ROSS1}, who exploited the separability of the scalar wave
equation in the electrovac C metric for this purpose. 

The Euclidean electrovac C metric possesses a hypersurface orthogonal
Killing vector, allowing analytic continuation to a well-defined real
Lorentzian metric in which scattering calculations may be
attempted. In a sense the use of the Lorentzian metric is nothing more
than a mathematical trick to compute scattering effects in the
Euclidean section. The Lorentzian section of the C metric contains a
pair of black holes accelerating apart, pulled by cosmic strings
extending to infinity. 

Hawking \& Ross argued that, although the quantum bubble metric need
not satisfy any field equations, its Lorentzian section would have a
structure like that of the C metric but without the axial
singularities. They further argued however that the C metric was a
sufficiently good model for the bubble metric in the scattering
calculation since the singularities would not adversely affect the
results. Their calculation seemed to confirm that, at least
semi-classically, quantum coherence of the incident scalar field is
lost.  

In anticipation of extending this latter calculation, the purpose of
this paper is to present, in a useful form, the classical perturbation
equations for incident fields of arbitrary spin (the Teukolsky 
equations~\cite{TEUK1}) in the C metric. After briefly reviewing some
properties of the C metric, -- in particular the `equal temperature
condition' and the `point-particle' limit relevant to virtual black
hole calculations, -- the full Teukolsky equations for arbitrary spin
fields are presented.  The separability of these equations into
one-dimensional `radial' and `angular' parts is then guaranteed (for 
all fields in the vacuum case and some fields in the 
electrovac case) due to the Petrov type D classification (Kamran 
\& McLenaghan~\cite{KM1}), and this separation is subsequently 
demonstrated. The resulting equations are then solved in the
`point-particle' limit to yield an approximate angular solution
together with an angular momentum quantisation condition, and an
approximate radial solution. Transmission and reflection coefficients
are then found for incident fields of a specific initial
configuration, following Hawking \& Ross.

\sect{Structure of the C Metric}

The electrovac C metric (\cite{KW1} and further references therein)
is most commonly expressed in the form
\be
ds^2 = \frac{1}{A^2\{x-y\}^2} \left\{ -p(y)dt^2 + \frac{dy^2}{p(y)} -
\frac{dx^2}{p(x)} - p(x)d\varphi^2 \right\}.
\label{metric}
\ee
In this notation the `metric polynomial' $p(\xi)$ is a quartic which
may be written as
\be
p(\xi) = \{ 1 + \eta \xi \} \{ 1 - \xi^2 - \zeta \xi^3 \}
\label{metpoly1}
\ee
where $\eta$ and $\zeta$ are dimensionless positive constants
constrained such that $p(\xi)$ has four real roots. 
This solution describes a pair of black holes of mass $M$ and charge
$\pm Q$ respectively, accelerating away from each other. The
parameters are related via
\be
QA = \pm \sqrt{\eta \zeta},
\ee
and
\be 
MA = \frac{1}{2} \{ \eta + \zeta \}.
\ee
The vacuum C metric solution may trivially be obtained from the above
by restricting $\eta \equiv 0$.

The coordinates $(t,y,x,\varphi)$ are suitably adapted to the timelike
Killing vector $\xi^a = A \delta_0^{\:a}$, the spacelike Killing
vector $\zeta^a = \delta_3^{\:a}$, and the nondegenerate eigenvector
$\eta^a = \delta_1^{\:a}$ of $R_{ab}^{(3)}$. It is thus clear from the
form of equation (\ref{metric}) that the solution is
static, time reversible, and axially symmetric, although these
coordinates cover only the neighbourhood of one black hole.

\begin{figure}
\begin{picture}(0,0)(0,0)
\put(22,82){\small $\xi_1$}
\put(137,82){\small $\xi_2$}
\put(241,82){\small $\xi_3$}
\put(358,82){\small $\xi_4$}
\put(402,74){\small $\xi$}
\put(316,214){\small $p(\xi)$}
\end{picture}   
\centering\epsfig{file=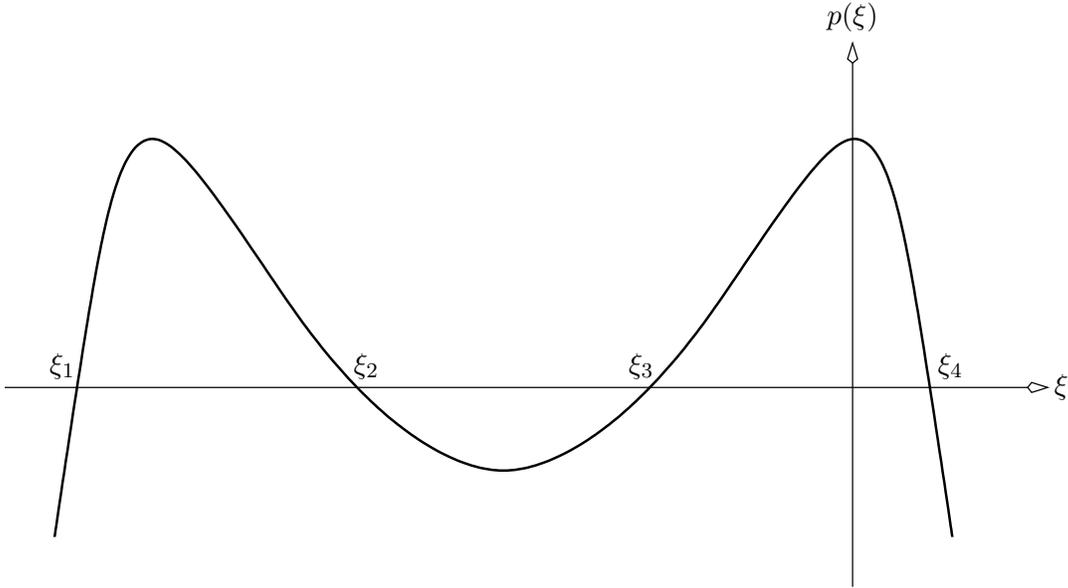,width=14cm}   
\caption{The metric polynomial $p(\xi)$, constrained with the `equal
temperature condition' (\ref{eqtemp1}).}
\label{quartic}
\end{figure}

If the roots of the metric polynomial are labelled $\xi_i$ which, for 
non-extreme solutions, must satisfy $\xi_1 < \xi_2 < \xi_3 < \xi_4$, then  
equation (\ref{metpoly1}) may be re-expressed as
\be
p(\xi) = -\eta \zeta \{\xi - \xi_1\}\{\xi - \xi_2\}\{\xi -
\xi_3\}\{\xi - \xi_4\}.
\label{metpoly2}
\ee    
To ensure the correct signature the coordinates must be restricted
such that $x \in [\xi_3,\xi_4]$ and $y \in (-\infty,x]$. The roots
then occur as shown in figure \ref{quartic} where the 
point $x = y = \xi_3$ is spatial infinity, with null and
timelike infinity along $x = y \neq \xi_3$.

Potential conical singularities exist along the axes $x = \xi_3$ and
$x = \xi_4$. Conventionally the singularity between the black holes
(along $\xi_4$) is removed by insisting periodicity in the azimuthal
coordinate such that
\be
\Delta \varphi = \frac{4 \pi}{| p'(\xi_4) |},
\label{phi_period}
\ee
which unavoidably produces a conical deficit along $x = \xi_3$. In the
C metric this is generally interpreted as a cosmic string lying along
the axis and terminating on the black hole. The deficit can be removed
by introducing an external gauge field~\cite{ERNST} -- the Ernst
solution, but the Lorentzian section of this metric is no longer
asymptotically flat.

The Euclidian section of the C metric may be obtained in the
conventional manner by setting $t = i \tau$ in equation
(\ref{metric}), where in addition the $y$ coordinate must be restricted
to the range $y \in [\xi_2,\xi_3]$ to ensure the metric is positive
definite. Further potential conical singularities then arise at $y =
\xi_2$ and $y = \xi_3$ which must both be eliminated in order that the
metric be regular. This may be achieved~\cite{ROSS2} by requiring
$\tau$ to be periodic, with 
\be
\Delta \tau = \frac{4 \pi}{p'(\xi_3)}
\ee
and by imposing an additional `equal temperature condition' on
$p(\xi)$ such that
\be
|p'(\xi_2)| = |p'(\xi_3)|.
\label{eqtemp1}
\ee
The latter condition (\ref{eqtemp1}), only realisable in the
electrovac case, may be re-expressed as a
relation between the parameters $\eta$ and $\zeta$, 
or alternatively as equivalent to insisting the roots $\xi_i$ of 
$p(\xi)$ satisfy
\be
\xi_2 - \xi_1 = \xi_4 - \xi_3.
\ee
With this regularity condition the topology of the Euclidean section 
becomes $S^2 \times S^2 - \{\mathrm{point}\}$, where the point removed 
is spatial infinity $x = y = \xi_3$. 

Of particular relevance to virtual black hole calculations is the
`point-particle' limit, in which the black holes are small on scales
set by the acceleration. Since an implicit relation holds between
$\eta$ and $\zeta$, there exists only one dimensionless parameter in
the metric, and so without loss of generality the `point-particle' 
limit may be expressed as
\be
\zeta \ll 1.
\label{pp_lim}
\ee 
In this limit deviations from spherical symmetry in the $(x, \varphi)$
section become small, as can be seen by making a coordinate
transformation $x \rightarrow \cos\theta$ so that $p(x) =
\sin^2\theta + \mathcal{O}(\zeta)$. This result evidently has
implications for the solution of any `angular' equations for incident
fields. 

\begin{figure}
\begin{picture}(0,0)(0,0)
\put(95,220){\scriptsize $\mathcal{J}^+$}
\put(161,220){\scriptsize $\mathcal{J}^+$}
\put(97,40){\scriptsize $\mathcal{J}^-$}
\put(155,40){\scriptsize $\mathcal{J}^-$}
\put(33,163){\scriptsize $H_{br}^+$}
\put(106,163){\scriptsize $H_{al}^+$}
\put(143,163){\scriptsize $H_{ar}^+$}
\put(215,163){\scriptsize $H_{bl}^+$}
\put(35,99){\scriptsize $H_{bl}^-$}
\put(105,99){\scriptsize $H_{ar}^-$}
\put(144,99){\scriptsize $H_{al}^-$}
\put(214,99){\scriptsize $H_{br}^-$}
\end{picture}   
\centering\epsfig{file=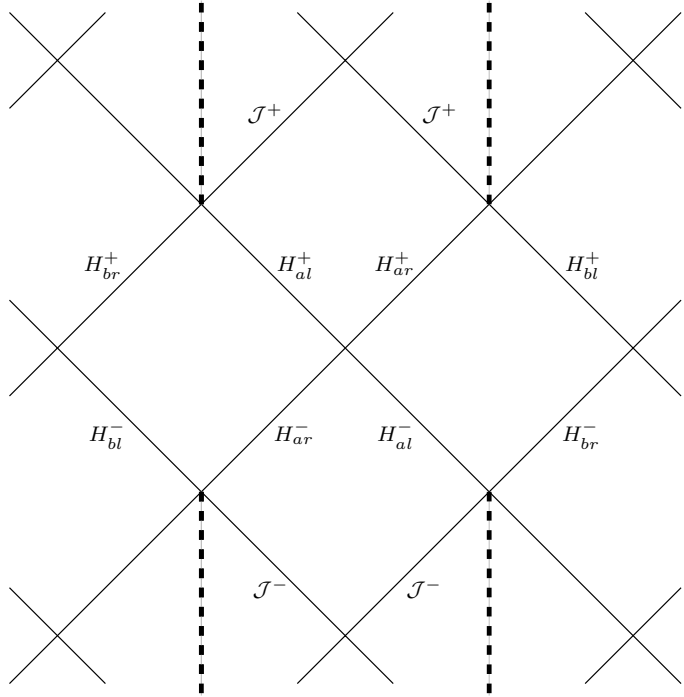,width=9cm}
\caption{The causal structure of the electrovac C metric in the
`point-particle' limit, neglecting the axis $x = \xi_3$. The notation
follows that of Hawking~\cite{SWH2} and Hawking \& Ross~\cite{ROSS1},
where $H_{bl}$, $H_{br}$, $H_{al}$ and $H_{ar}$ are the left and right
acceleration and outer black hole horizons respectively. The future
halves of each are denoted $+$ and the past halves $-$. The heavy
dashed lines are curvature singularities at `$r=0$'.}
\label{CAUSAL}
\end{figure}

Equation (\ref{metpoly1}) may be solved for the roots $\xi_i$ which,
under the assumption (\ref{pp_lim}), can be written as series in
$\zeta$ to give
\ba
\xi_1 & = & -\frac{1}{\zeta} - 2 + \zeta + \mathcal{O}(\zeta^2), \\
\xi_2 & = & -\frac{1}{\zeta} + \zeta + \mathcal{O}(\zeta^2), \\
\xi_3 & = & -1 - \frac{1}{2} \zeta + \mathcal{O}(\zeta^2),  \\
\xi_4 & = & 1 - \frac{1}{2} \zeta + \mathcal{O}(\zeta^2).
\ea
Clearly then, the acceleration horizon at $y = \xi_3$ is largely
unaffected by variations in the parameter $\zeta$ if (\ref{pp_lim})
holds, whereas the inner and outer black hole horizons $y = \xi_1$ and
$\xi_2$ respectively are both highly sensitive. As $\zeta \rightarrow
0$ these roots $\rightarrow - \infty$ in unison which, together with
the identification of infinity at $x = y$, supports the definition of a
radial coordinate $r$~\cite{KW1} as
\be
r^{-1} = A \{x-y\}.
\label{r_coord}
\ee
If the axis $x = \xi_3$ is neglected all observers will intersect the
acceleration horizon before reaching infinity and, in the
`point-particle' limit, the causal structure approaches that shown in
figure~\ref{CAUSAL}. The surface gravity $\kappa$ at the outer black
hole horizon and the acceleration horizon, is given by
\be
\kappa \equiv \frac{2\pi}{\Delta \tau} = \frac{p'(\xi_3)}{2} = 1 - 2
\zeta + \mathcal{O}(\zeta^2). 
\label{s_grav}
\ee

Quite apart from the additional conditions placed on the parameters to
remove conical singularities and render the Euclidean section regular,
the C metric possesses a great deal of intrinsic symmetry which may be
suitably exploited in the analysis of incident perturbing fields. In
particular, the appropriately normalised null linear combinations of 
$\xi^a$ and $\eta^a$ are the repeated principal null vectors of both 
the Weyl tensor $C_{abcd}$~\cite{KW1} and the Maxwell tensor 
$F_{ab}$~\cite{DKM}. The existence of these repeated null vectors
implies the metric is of Petrov type D, and perturbing fields may
therefore be treated using the methods first developed by Teukolsky 
for the Kerr solution.

\sect{Derivation of the Teukolsky Equation}

The C metric and its associated electrovac generalisation are examples
of algebraically special spacetimes, -- in particular they are of
Petrov type $\{2,2\}$ or D which amounts to saying that the Weyl
spinor $\Psi_{ABCD}$ may be written as
\be
\Psi_{ABCD} = o_{(A} o_B \iota_C \iota_{D)}
\ee
for a suitably chosen spin basis $(o,\iota)$. This is
equivalent~\cite{DKM} to the existence of two real null vector fields
$l$ and $n$, everywhere satisfying 
\be
l^b l^c C_{abc[d} l_{e]} = n^b n^c C_{abc[d} n_{e]} = 0.
\ee
$l$ and $n$ are the repeated principal null vectors of the Weyl tensor
$C_{abcd}$ which, for the electrovac solution, coincide with those for
the Maxwell tensor so that
\be
l^a F_{a[b} l_{c]} = n^a F_{a[b} n_{c]} = 0.
\ee
For the type D solutions the principal null congruences defined
respectively by $l$ and $n$ are both geodesic and shear-free. These
conditions are sufficient to guarantee, via the generalised
Goldberg-Sachs theorem~\cite{JS1}, the existence of a canonical null
tetrad $(l, n, m, \overline{m})$ in which the Newman-Penrose spin
coefficients $\kappa, \sigma, \lambda,$ and $\nu$ vanish identically,
and in which the only non-zero tetrad components of the Weyl and Ricci
spinors are $\Psi_2$ and $\Phi_{11}$ respectively. In the case of 
vacuum solutions, $\Phi_{11}$ is also identically zero.

Assuming these results in a vacuum background, Teukolsky~\cite{TEUK1}
combined the Newman-Penrose form of the linearised Ricci and Bianchi 
identities in pairs to generate decoupled equations satisfied by the 
first order gravitational perturbations $\Psi_0$ and $\Psi_4$. By
similarly combining the Newman-Penrose form of the Maxwell identities
he also found decoupled equations for the tetrad components $\phi_0$ and
$\phi_2$ of a test Maxwell field, and likewise for the tetrad
components $\chi_0$ and $\chi_1$ of a test neutrino field. Introducing
a parameter $s$ as the spin-weight~\cite{GHP1} of the field 
components allows the homogeneous form of these equations to be 
combined and written as
\ba
\lefteqn{\Bigl\{[D-(2s-1)\epsilon+\overline{\epsilon}-2s\varrho- 
\overline{\varrho}](\Delta-2s\gamma+\mu)\Bigr.} \nonumber \\[0.8mm] 
& & -[\delta-(2s-1)\beta+\overline{\pi}-2s\tau-\overline{\alpha}]
(\overline{\delta}+\pi-2s\alpha)  \nonumber \\[0.8mm]
& & \mbox{\hspace{10mm}} -\Bigl.(2s-1)(s-1)\Psi_2\Bigr\}\Phi_s = 0 
\label{SPIN_PLUS}
\ea
for $s>0$, and
\ba
\lefteqn{\Bigl\{[\Delta-(2s+1)\gamma-\overline{\gamma}-2s\mu+
\overline{\mu}](D-2s\epsilon-\varrho)\Bigr.} \nonumber \\[0.8mm]
& & -[\overline{\delta}-(2s+1)\alpha-\overline{\tau}-2s\pi+ 
\overline{\beta}](\delta-2s\beta-\tau) \nonumber  \\[0.8mm]
& & \mbox{\hspace{10mm}} -\Bigl.(2s+1)(s+1)\Psi_2\Bigr\}\Phi_s = 0
\label{SPIN_MINUS}
\ea
for $s<0$. The $\Phi_s$'s are to be interpreted as the tetrad
components of the appropriate field, which are conventionally defined
as $\Psi_4$, $\phi_2$ and $\chi_1$ for spin-weights -2, -1, and
-$\frac{1}{2}$, and $\Psi_0$, $\phi_0$ and $\chi_0$ for spin-weights
2, 1, and $\frac{1}{2}$ respectively.

The above results then also hold for type D electrovac solutions
(where $\Phi_{11} \neq 0$) with the exception of $s = \pm 2$, under
the assumption that the spin 1 perturbations are not in the
background Maxwell field but in another gauge field.  

In order to proceed, a choice of tetrad basis $(l, n, m,
\overline{m})$ for the C metric must be made. The conventional reality
condition on the basis vectors $l$ and $n$ requires that a different 
choice of tetrad basis be made for the regions in which the metric 
polynomial $p(y)$ is greater than or less than zero. Dealing first 
with those regions in which $p(y) > 0$, a suitable choice is given by
\be
\left. \begin{array}{ccr}
l_a & = & [\sqrt{p(y)},1/\sqrt{p(y)}, 0, 0] \\[0.8mm]
n_a & = & [-\sqrt{p(y)},1/\sqrt{p(y)}, 0, 0] \\[0.8mm]
m_a & = & [0, 0, i/\sqrt{p(x)}, -\sqrt{p(x)}] \\
\end{array} \right\} \times \frac{1}{\sqrt{2}A\{x-y\}}.
\ee
This is the canonical symmetric tetrad, in terms of which the non-zero
basis components of the Weyl and Ricci  
spinors are 
\be
\Psi_2 = \frac{1}{12} A^2 \{ x-y \}^2 [ p''(x) - p''(y) ]
\label{psi2}
\ee
and
\ba
\lefteqn{\Phi_{11}=-\frac{1}{8}A^2\Bigl(\{x-y\}^2[p''(x)+p''(y)] \Bigr.}
\nonumber \\
&&\mbox{\hspace{20mm}}\Bigl.-2\{x-y\}[p'(x)-p'(y)] \Bigr) 
\ea
respectively. The inherent symmetries of the chosen tetrad force the 
eight remaining non-zero Newman-Penrose spin coefficients into two 
real and two complex pairs, so that
\ba
-\varrho = \mu & = & \frac{1}{\sqrt{2}} A \sqrt{p(y)}, \\
-\gamma = \epsilon & = & \frac{1}{\sqrt{2p(y)}} A \left\{
\frac{1}{2}\{x-y\} p'(y) + p(y) \right\}, \\
\tau = \mbox{}\pi & = & \frac{i}{\sqrt{2}} A \sqrt{p(x)}, \\
\alpha = \mbox{}\beta & = & \frac{i}{\sqrt{2p(x)}} A \left\{
\frac{1}{2}\{x-y\} p'(x) - p(x) \right\}.
\ea
Finally, the four directional derivative operators $D$, $\Delta$,
$\delta$, and $\overline{\delta}$ become
\ba
D & = & \frac{1}{\sqrt{2}}A\{x-y\}\left(\sqrt{p(y)}\frac{\partial}
{\partial y}  - \frac{\partial}{\partial t} \right), \\
\Delta & = & \frac{1}{\sqrt{2}}A\{x-y\}\left(\sqrt{p(y)}\frac
{\partial}{\partial y} + \frac{\partial}{\partial t} \right), \\
\delta &=&\frac{1}{\sqrt{2}}A\{x-y\}\left(-i\sqrt{p(x)}\frac
{\partial}{\partial x} + \frac{\partial}{\partial \varphi} \right), \\
\overline{\delta}&=&\frac{1}{\sqrt{2}}A\{x-y\}\left( i\sqrt{p(x)}
\frac{\partial}{\partial x} + \frac{\partial}{\partial \varphi}
\right). 
\label{delta_bar} 
\ea

Substituting the relations (\ref{psi2} - \ref{delta_bar}) into equation
(\ref{SPIN_PLUS}), under the assumption that $\Phi_s$ is functionally
dependent on all coordinates, results in a second order partial
differential equation for positive spin-weight fields. This may be
combined with the result obtained from a similar substitution into
(\ref{SPIN_MINUS}) and written as
\ba
\lefteqn{ A^2 \Biggl( \{x-y\}^2 \Biggl[ \frac{1}{2}\partial_y \{
p(y)\partial_y \} - \frac{1}{2p(y)} \left\{ \partial_t +\frac{1}{2}
sp'(y) \right\}^2  - \frac{1}{2}\partial_x \{p(x)\partial_x \} - 
\frac{1}{2p(x)} \left\{ \partial_{\varphi} + \frac{i}{2}sp'(x)
\right\}^2 \Biggr. \Biggr. } \nonumber \\[0.8mm]
& & \Biggl. -\frac{1}{12}(2s^2+1) \{p''(x)-p''(y)\} \Biggr] + \{x-y\} 
(|s|+1) \left\{ p(y)\partial_y + \frac{1}{2}p'(y) + p(x)\partial_x +
\frac{1}{2}p'(x) \right\} \nonumber \\[0.8mm] 
& & \Biggl. \mbox{\hspace{10mm}} - \frac{1}{2}(s^2+3|s|+2)\{p(x)-p(y)\} 
\Biggr) \Phi_s = 0 
\label{TEUK_EQ}
\ea 
which is then valid for massless fields of spin-weight $s = \pm
\frac{1}{2}$, $\pm 1$ and $\pm 2$. 

For those regions of the C metric in which $p(y)<0$, a suitable choice
of tetrad may be obtained from the original with a boost of 
$-i$ under which the Newman-Penrose quantities $\Psi_2$ and
$\Phi_{11}$ remain unchanged, the non-zero spin coefficients transform
as  
\be
\begin{array}{ccccccc}
\tilde{\mu} & = & i \mu, & \mbox{\hspace{1cm}} & \tilde{\varrho} & = & 
-i\varrho, \\ \tilde{\epsilon} & = & -i \epsilon, & \mbox{\hspace{1cm}}
 & \tilde{\gamma} & = & i \gamma, \\ \tilde{\tau} & = & -\tau, & 
\mbox{\hspace{1cm}} & \tilde{\pi} & = & -\pi, \\ \tilde{\alpha} & = 
& -\alpha, & \mbox{\hspace{1cm}} & \tilde{\beta} & = & -\beta, 
\end{array}
\ee
and $\Phi_s$ acquires a phase factor
\be
\Phi_s \rightarrow \tilde{\Phi}_s = (-i)^s \Phi_s.
\ee
Substituting these new relations as before produces identical results
to those for the $p(y)>0$ regions, and so the Teukolsky equation
(\ref{TEUK_EQ}) is valid throughout the C metric solution. Also, by
setting $s=0$, the expression reduces to the C metric form of the
conformally invariant massless scalar equation   
\be
\left\{ \Box - \frac{1}{6} R \right\} \Phi_0 = 0,
\ee 
and so the domain of definition may be extended to include
null scalar fields. Equation (\ref{TEUK_EQ}) may therefore be
interpreted as describing massless fields of spin-weight $s = 0$,
$\pm\frac{1}{2}$, $\pm 1$ and $\pm 2$ incident upon the vacuum
C metric. In addition, it remains valid in the electrovac C
metric when restricted to massless test fields of spin-weight $0$, $\pm
\frac{1}{2}$ and $\pm 1$. 

The form of equation (\ref{TEUK_EQ}) again highlights the inherent
symmetries of the chosen tetrad which effectively produce terms in
matched pairs -- one in $x$ and one in $y$. It is hardly surprising
therefore that, with a suitably chosen ansatz for $\Phi_s$, the
equation separates into two ordinary differential
equations in the coordinates $x$ and $y$ respectively. Indeed, the
algebraically special character of the C metric was suitably exploited
in choosing an appropriate tetrad basis for this exact purpose. This
separation procedure has been shown to be possible, given a suitably
chosen basis, for all type D vacuum metrics by Kamran \&
McLenaghan~\cite{KM1}, and will be explicitly demonstrated for the C
metric in the next section.  

\sect{Separation of Variables}

Having derived the expressions for arbitrary spin massless perturbations
in type D vacuum solutions, Teukolsky demonstrated~\cite{TEUK1} that
for the Kerr metric the equations separated when expressed in a
Kinnersley~\cite{KINN} tetrad. In view of the validity of Teukolsky's
methods, Kamran \& McLenaghan conjectured that a similar separation
could be performed for the entire class of type D vacuum
spacetimes. This was subsequently demonstrated for $s=0$,
$\pm\frac{1}{2}$, $\pm1$ and $\pm2$ in~\cite{KM1}. 

For the C metric, the nature of the conformal factor $A\{x-y\}$~\cite{BD1}
and the ignorable coordinates $t$ and $\varphi$ in the metric strongly
suggest an ansatz of the form
\be
\Phi_s = [ A \{x-y\} ]^{|s|+1} \mathrm{e}^{ -i(\omega t -
m \varphi)} \Theta_s(x,y).
\label{ansatz}
\ee
Since $\varphi$ is a periodic coordinate (\ref{phi_period}), $m$ must
be of the form
\be
m = m_0 \frac{|p'(\xi_4)|}{2} = m_0 \left\{ 1 + 2\zeta +
\mathcal{O}(\zeta^2) \right\}
\ee
where $m_0$ is, without loss of generality, a positive integer.
 
For the $s > 0$ fields, acting with the first two operators in
the Newman-Penrose form of the Teukolsky equation (\ref{SPIN_PLUS})
yields  
\be
\Bigl\{\Delta - 2s\gamma + \mu \Bigr\}\Phi_s =
\frac{1}{\sqrt{2}}[A\{x-y\}]^{s+2}\mathrm{e}^{-i(\omega t - m\varphi
)} \mathcal{L}_{ys_+} \Theta_s(x,y),
\ee
and
\be
\Bigl\{\overline{\delta} - 2s\alpha + \pi\Bigr\} \Phi_s = 
\frac{1}{\sqrt{2}}[A\{x-y\}]^{s+2}\mathrm{e}^{-i(\omega t - m\varphi 
)} \mathcal{L}_{xs_+} \Theta_s(x,y),
\ee
where the differential operators $\mathcal{L}_{ys_+}$ and 
$\mathcal{L}_{xs_+}$ may be written as
\ba
\mathcal{L}_{ys_+} & = & \sqrt{p(y)}\frac{\partial}{\partial y} -
\frac{1}{\sqrt{p(y)}} \left\{ i\omega - \frac{1}{2} s p'(y)
\right\}, \\
\mathcal{L}_{xs_+} & = & i \sqrt{p(x)}\frac{\partial}{\partial x} +
\frac{i}{\sqrt{p(x)}} \left\{ m + \frac{1}{2} s p'(x)
\right\}.  
\ea
Similarly for the $s < 0$ fields, acting on the same ansatz with the 
first two operators in equation (\ref{SPIN_MINUS}) gives
\be
\Bigl\{D - 2s\epsilon - \varrho \Bigr\}\Phi_s =
\frac{1}{\sqrt{2}}[A\{x-y\}]^{-s+2}\mathrm{e}^{-i(\omega t - m\varphi 
)} \mathcal{L}_{ys_-} \Theta_s(x,y),
\ee
and
\be
\Bigl\{\delta - 2s\beta - \tau\Bigr\} \Phi_s = 
\frac{1}{\sqrt{2}}[A\{x-y\}]^{-s+2}\mathrm{e}^{-i(\omega t - m\varphi 
)} \mathcal{L}_{xs_-} \Theta_s(x,y), 
\ee
where now the operators $\mathcal{L}_{ys_-}$ and 
$\mathcal{L}_{xs_-}$ take the form
\ba
\mathcal{L}_{ys_-} & = & \sqrt{p(y)}\frac{\partial}{\partial y} +
\frac{1}{\sqrt{p(y)}} \left\{ i\omega - \frac{1}{2} s p'(y)
\right\}, \\
\mathcal{L}_{xs_-} & = & -i \sqrt{p(x)}\frac{\partial}{\partial x} +
\frac{i}{\sqrt{p(x)}} \left\{ m + \frac{1}{2} s p'(x)
\right\}.  
\ea
These two sets of equations may obviously be combined into a single
pair valid for $s = 0$, $\pm \frac{1}{2}$, $\pm 1$, and $\pm 2$ by
defining 
\be
\mathcal{L}_{ys} \stackrel{\mathrm{def}}{=} \sqrt{p(y)}\frac{\partial}
{\partial y} - \frac{|s|}{s} \frac{1}{\sqrt{p(y)}} \left\{ i \omega -
\frac{1}{2} s p'(y) \right\},
\ee
and
\be
\mathcal{L}_{xs} \stackrel{\mathrm{def}}{=} i \frac{|s|}{s}
\sqrt{p(x)} \frac{\partial}{\partial x} + \frac{i}{\sqrt{p(y)}}
\left\{ m + \frac{1}{2} s p'(x) \right\}.
\ee

The remaining operators in the Newman-Penrose form of the Teukolsky
equations (\ref{SPIN_PLUS}, \ref{SPIN_MINUS}) may be similarly expanded
by allowing them to act on a coordinate-dependant function
$\Lambda_s(t,y,x,\varphi)$ of the form 
\be
\Lambda_s(t,y,x,\varphi) = \mathrm{e}^{-i(\omega t - m \varphi)} \times
\Omega_s(x,y). 
\ee  
For the $s > 0$ case this leads to 
\ba
\lefteqn{ \Bigl\{ D - (2s-1)\epsilon + \overline{\epsilon} - 2s\varrho
- \overline{\varrho} \Bigr\} \Lambda_s = } \nonumber \\
& & \mbox{\hspace{20mm}}\mathrm{e}^{-i(\omega t - m \varphi)} \left\{
\frac{1}{\sqrt{2}}A\{x-y\} \mathcal{D}_{ys_+} + \mathcal{K}_{ys_+}
\right\} \Omega_s(x,y), \\[0.8mm] 
\lefteqn{\Bigl\{ \delta - (2s-1)\beta + \overline{\pi} - 2s\tau -
\overline{\alpha} \Bigr\} \Lambda_s = } \nonumber \\
& & \mbox{\hspace{20mm}}\mathrm{e}^{-i(\omega t - m \varphi)} \left\{
\frac{1}{\sqrt{2}} A \{x-y\} \mathcal{D}_{xs_+} + \mathcal{K}_{xs_+}
\right\} \Omega_s(x,y), 
\ea
and for the $s < 0$ case
\ba
\lefteqn{ \Bigl\{\Delta - (2s+1)\gamma + \overline{\gamma} - 2s\mu -
\overline{\mu} \Bigr\} \Lambda_s = } \nonumber \\
& & \mbox{\hspace{20mm}}\mathrm{e}^{-i(\omega t - m \varphi)} \left\{
\frac{1} {\sqrt{2}}A\{x-y\} \mathcal{D}_{ys_-} + \mathcal{K}_{ys_-}
\right\} \Omega_s(x,y), \\[0.8mm]
\lefteqn{ \Bigl\{\overline{\delta} - (2s+1)\alpha + \overline{\tau} 
- 2s\pi - \overline{\beta} \Bigr\} \Lambda_s = } \nonumber \\
& & \mbox{\hspace{20mm}}\mathrm{e}^{-i(\omega t - m \varphi)} \left\{ 
\frac{1}{\sqrt{2}} A \{x-y\}\mathcal{D}_{xs_-} + \mathcal{K}_{xs_-}
\right\} \Omega_s(x,y). 
\ea
As for the $\mathcal{L}$ operators, the $\mathcal{D}$'s and
$\mathcal{K}$'s may be combined into pairs defined for both
positive and negative $s$ which may be written as
\ba
\mathcal{D}_{ys} & \stackrel{\mathrm{def}}{=} & \sqrt{p(y)}
\frac{\partial}{\partial y}  + \frac{|s|}{s} \frac{1}{\sqrt{p(y)}}
\left\{ i \omega - \frac{1}{2}\left(s - \frac{|s|}{s}\right) p'(y)
\right\}, \\[0.8mm]
\mathcal{D}_{xs} & \stackrel{\mathrm{def}}{=} & -i \frac{|s|}{s} \sqrt{p(x)}
\frac{\partial}{\partial x}  + \frac{i}{\sqrt{p(x)}}
\left\{ m + \frac{1}{2}\left(s - \frac{|s|}{s}\right) p'(x) \right\},
\ea
and
\ba
\mathcal{K}_{ys} & \stackrel{\mathrm{def}}{=} & \frac{1}{\sqrt{2}}
\frac{|s|}{s} A \sqrt{p(y)} \left\{ s + \frac{2|s|}{s} \right\}, 
\\[0.8mm]
\mathcal{K}_{xs} & \stackrel{\mathrm{def}}{=} & \frac{i}{\sqrt{2}}
A \sqrt{p(x)} \left\{ s + \frac{2|s|}{s} \right\}. 
\ea

It should be noted at this stage that each of the $\mathcal{L}$'s,
$\mathcal{D}$'s and $\mathcal{K}$'s are functions of only one
coordinate, -- either $x$ or $y$. If these results are combined
appropriately then, given the form of the ansatz (\ref{ansatz}), the
Teukolsky equations (\ref{SPIN_PLUS}) and (\ref{SPIN_MINUS}) are seen
to be wholly equivalent to the single expression
\ba
\lefteqn{ \Bigg( \left\{ \frac{1}{\sqrt{2}}A\{x-y\}\mathcal{D}_{ys} +
\mathcal{K}_{ys} \right\} \left[\frac{1}{\sqrt{2}}[A\{x-y\}
]^{|s|+2} \mathcal{L}_{ys} \right] \Biggr. } \nonumber \\[0.8mm]
& &- \left\{ \frac{1}{\sqrt{2}}A\{x-y\}\mathcal{D}_{xs} +
\mathcal{K}_{xs} \right\} \left[\frac{1}{\sqrt{2}}[A\{x-y\}
]^{|s|+2} \mathcal{L}_{xs} \right] \nonumber \\[0.8mm]
& & \Biggl. \mbox{\hspace{15mm}}-(2|s|-1)(|s|-1)[A\{x-y\}] 
^{|s|+1}\Psi_2 \Biggr) \Theta_s(x,y) = 0.
\ea 
Considering the action of both $\mathcal{D}_{ys}$ and
$\mathcal{D}_{xs}$ on the term $[A\{x-y\}]^{|s|+2}$ and the explicit
form of $\Psi_2$ (\ref{psi2}) this may be substantially simplified,
becoming 
\ba
\lefteqn{ \Bigl( \{ \mathcal{D}_{ys} \mathcal{L}_{ys} + \frac{1}{6}(2s^2
- 3|s| + 1) p''(y) \} \ \Bigr. } \nonumber \\
& & \Bigl. \mbox{\hspace{5mm}} - \{ \mathcal{D}_{xs} \mathcal{L}_{xs}
+ \frac{1}{6}(2s^2 - 3|s| + 1)p''(x) \} \Bigr) \Theta_s(x,y) =  0
\ea
which is clearly separable under the assumption $\Theta_s(x,y) =
\phi_s(x) \times \psi_s(y)$. By expanding the $\mathcal{D}$ and
$\mathcal{L}$ operators, the separated expression for the
$y$-coordinate may be written as 
\be 
\left( \frac{\partial}{\partial y} \left\{ p(y)\frac{\partial}
{\partial y} \right\} + \frac{1}{p(y)}\left\{ \omega + \frac{i}{2} s
p'(y) \right\}^2 + \frac{1}{6}(2s^2+1)p''(y) + \Gamma_s \right)
\psi_s(y) = 0 
\label{Y_EQ}
\ee
where $\Gamma_s$ is a separation constant which will in general depend
on the spin-weight $s$. Likewise for the $x$-coordinate, the separated
equation becomes 
\be
\left( \frac{\partial}{\partial x} \left\{ p(x)\frac{\partial}
{\partial x}\right\} - \frac{1}{p(x)}\left\{ m + \frac{1}{2} s p'(x)
\right\}^2 + \frac{1}{6}(2s^2+1)p''(x) + \Gamma_s \right) \phi_s(x)  =
0. 
\label{X_EQ}
\ee

The pair of separated equations (\ref{Y_EQ}) and (\ref{X_EQ}) may be
said to describe the $y$ and $x$ components respectively of fields
with all spin-weights considered incident on the vacuum C metric. As
before, they continue to hold in the electrovac C metric if restricted
to massless test fields of spin-weight $s=0$, $\pm \frac{1}{2}$ and
$\pm 1$. 

The unification and subsequent separation of the equations describing
incident fields of a variety of spin-weights evidently represents an
appreciable success for the Newman-Penrose formalism. However the
resulting expressions, when considered as a pair of second order
ordinary differential equations, are seen to possess five
regular singular points -- the four roots of $p(\xi)$ and $\infty$ --
and hence cannot be solved analytically. In order to furnish even an
approximate solution, the `point-particle' limit must be exploited to
suitably simplify both equations.

\sect{Solving the \mbox{\boldmath $x$} Equation}

Equation (\ref{X_EQ}) for $\phi_s(x)$, together with suitable
conditions of regularity at the boundaries $x = \xi_3$ and $\xi_4$
represents a Sturm-Liouville eigenvalue problem for the separation
constant $\Gamma_s$. For fixed $s$ and $m$ the eigenvalues may be
labelled by $j$, where it is assumed that 
\be
j = j_0 + \zeta j_1 + \mathcal{O}(\zeta^2).
\ee
Since the expression is self-adjoint, its eigenfunctions -- labelled
by $s$, $m$ and $j$, -- will be both complete and orthogonal on the
interval $x \in [\xi_3,\xi_4]$, but the existence of five regular 
singular points implies these solutions will not have a closed form in
terms of known functions. However, considerable analytic progress may
be made by moving to the `point-particle' limit after first changing
variables 
\be
x \rightarrow \hat{x} = \frac{1}{\beta}\left(x - \frac{1}{2}\{\xi_3 +
\xi_4\} \right)
\ee
where
\be 
\beta \stackrel{\mathrm{def}}{=} \frac{1}{2}\{\xi_4 - \xi_3\}.
\ee
The range of $x$ is then mapped to $\hat{x} \in [-1,1]$, and the
metric polynomial may be re-written as
\be
p(\hat{x}) = -\eta \zeta \beta^4 \Bigl\{(\hat{x} +\alpha)^2 -
1\Bigr\} \Bigl\{ \hat{x}^2 - 1\Bigr\}
\ee
with 
\be
\alpha \stackrel{\mathrm{def}}{=} \frac{1}{\beta}\left\{ \xi_3 - \xi_1
\right\}.
\ee
Having made this change, equation (\ref{X_EQ}) becomes
\be
\left( \frac{1}{\beta^2}\frac{\partial}{\partial \hat{x}} \left\{
p(\hat{x}) \frac{\partial}{\partial \hat{x}} \right\} -
\frac{1}{p(\hat{x})} \left\{ m + \frac{1}{2} s p'(\hat{x})
\right\}^2 + \frac{1}{6}(2s^2+1)p''(\hat{x}) + \Gamma_s \right)
{{}_s\phi_{jm}}(\hat{x}) = 0
\label{NEWX_EQ}
\ee
where $'$ denotes differentiation with respect to $x$, and
$\partial_{\hat{x}} = \beta \partial_x$.

In the limit $\zeta \ll 1$ the three polynomials $p(\hat{x})$,
$p'(\hat{x})$ and $p''(\hat{x})$ may be expressed as power
series in $\zeta$
\ba
p(\hat{x}) & = & (1-\hat{x}^2) \bigg\{ 1 + 2\hat{x}\zeta +
\mathcal{O}(\zeta^2) \bigg\}, \\[0.8mm]
p'(\hat{x}) & = & -2 \bigg\{ \hat{x} - (1-3\hat{x}^2)\zeta +
\mathcal{O}(\zeta^2) \bigg\}, \\[0.8mm]
p''(\hat{x}) & = & -2 \bigg\{ 1 + 6\hat{x}\zeta + \mathcal{O}(\zeta^2)
\bigg\}.
\ea
If it is further assumed that the eigenfunctions may also be written
as a power series in $\zeta$ 
\be
{{}_s^{\phantom{0}} \phi_{jm}^{\phantom{0}}} =
{{}_s^{\phantom{0}} \phi_{jm}^{(0)}} + {{}_s^{\phantom{1}}
\phi_{jm}^{(1)}} \zeta + \mathcal{O}(\zeta^2) 
\ee
then (\ref{NEWX_EQ}) may be separated into a series of equations for
these individual functions, the first of which is
\be
\left( \frac{\partial}{\partial \hat{x}} \left\{ (1-\hat{x}^2)
\frac{\partial}{\partial \hat{x}} \right\} - \frac{ \left\{ m_0 - s
\hat{x} \right\}^2 } {1 - \hat{x}^2} - \frac{1}{3}(2s^2+1) + \Gamma_s
\right) {{}_s^{\phantom{0}}\phi_{jm}^{(0)}}(\hat{x}) = 0.
\ee
If the label $j$ is defined via the equation 
\be
\Gamma_s = j(j+1) - \frac{1}{3}(s^2 - 1)
\label{gamma_s}
\ee
then, with an additional change of variables $\hat{x} \rightarrow -
\cos \theta$, this reduces to
\be
\left( \frac{1}{\sin \theta} \frac{\partial}{\partial \theta} \left\{
\sin \theta \frac{\partial}{\partial \theta} \right\} - \left\{
\frac{m_0 + s \cos \theta}{\sin \theta} \right\}^2 - s^2 + j_0(j_0+1)
\right) {{}_s^{\phantom{0}}\phi_{jm}^{(0)}}(\theta) = 0.
\ee
Since the zeroth order in $\zeta$ corresponds to a limit of spherical
symmetry in the $(\theta, \varphi)$ section it is hardly surprising
that the solutions are related to the spin-weighted spherical
harmonics~\cite{GMNRS} 
\be
{{}_s^{\phantom{0}}Y_{jm}^{ \phantom{0}}}(\theta, \varphi) =
{{}_s^{\phantom{0}}\phi_{jm}^{(0)}}(\theta) \times \mathrm{e}^{i m
\varphi}.
\ee
$j_0$ must therefore be half-integer, satisfying $j_0
\ge \mathrm{max}(|m_0|, |s|)$ -- so in this limit $j_0$ and $m_0$ are
the usual `total angular momentum' and `angular momentum about the 
symmetry axis' quantum numbers respectively. 

Strictly this derivation has assumed the relation implicit between the
parameters $\eta$ and $\zeta$ of the metric due to the imposed `equal
temperature condition' (\ref{eqtemp1}). In this sense the solutions to
(\ref{X_EQ}), expressed in the `point-particle' limit as
\be
\phi_s(x) = {}_s\phi_{jm}(\theta) = \mathrm{e}^{-i m \varphi} \times
{}_s Y_{jm}(\theta, \varphi) + \mathcal{O}(\zeta)
\label{x_sol}
\ee
describe the $x$ component of massless test fields of spin-weight $s =
0$, $\pm \frac{1}{2}$ and $\pm 1$ only, incident on the electrovac C
metric. This is the desired result for virtual black hole
calculations, but for the sake of completeness, to zeroth order in
$\zeta$ the metric polynomial and its derivatives are identical in the
vacuum ($\eta \equiv 0$) case, and so (\ref{x_sol}) is in addition
valid for all spin-weights incident on the vacuum C metric.

\sect{Boundary Conditions on the \mbox{\boldmath $y$} Equation.}

The evolution of fields with spin-weight $s$ described by equation
(\ref{Y_EQ}) essentially defines a one-dimensional scattering problem,
upon which suitable asymptotic boundary conditions must be
imposed in order that the problem be well-defined. Following
Hawking \& Ross~\cite{ROSS1}, initial data for the incident fields may
be defined on an `initial' Cauchy surface suitably constructed from
the left acceleration and right black hole horizons. The scattering
problem then becomes the propagation of this data forward through the
spacetime to a `future' Cauchy surface constructed from the future
halves of the left and right black hole horizons and
$\mathcal{J}^+$. The form of the initial data on the acceleration and
black hole horizons is nonetheless restricted by the behaviour of the
fields as the metric polynomial $p(y) \rightarrow 0$, since it is in
this limit that the horizons are approached.  

With the exception of fields incident along the axis $x = \xi_3$ this
problem may conveniently be divided into two sections -- propagation
from $H_{al}^-$ and $H_{br}^-$ to $H_{ar}^+$ and $H_{bl}^+$, followed
by propagation from $H_{ar}^+$ and $H_{al}^+$ to $\mathcal{J}^+$. The
axially incident portion of the field is an exceptional case since
this is the unique component which reaches $\mathcal{J}^+$ before
intersecting the acceleration horizon. However this contribution may
consistently be disregarded as a set of measure zero amongst the
entirety of the incident field.

\begin{figure}
\begin{picture}(0,0)(0,0)
\put(-215,-85){\scriptsize $H_{br}^+$}
\put(-158,-85){\scriptsize $H_{al}^+$}
\put(-109,-135){\scriptsize $H_{al}^-$}
\put(-78,-135){\scriptsize $H_{br}^-$}
\put(-193,-103){\scriptsize $u$}
\put(-169,-103){\scriptsize $v$}
\put(50,-85){\scriptsize $H_{br}^+$}
\put(103,-38){\scriptsize $\mathcal{J}^+$}
\put(155,-38){\scriptsize $\mathcal{J}^+$}
\put(200,-85){\scriptsize $H_{bl}^+$}
\put(165,-131){\scriptsize $v$}
\put(188,-131){\scriptsize $u$}
\end{picture}   
\centering\epsfig{file=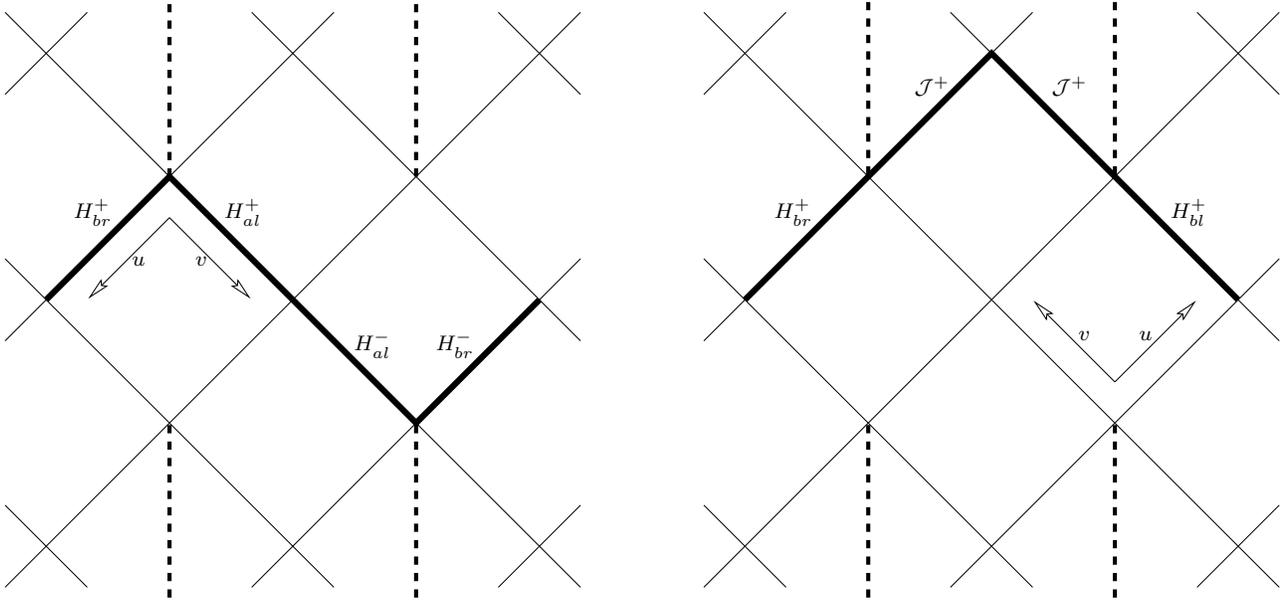,width=17cm}
\caption{Penrose diagrams for the electrovac C metric showing the
initial and future Cauchy surfaces respectively, and the null
coordinates $u,v = t \mp y_*$ in two `Rindler diamonds'.}
\label{hor_diag}
\end{figure}

Further insight into the asymptotic behaviour of equation (\ref{Y_EQ})
may be gained by defining a conventional `tortoise' coordinate $y_*$
such that 
\be
dy_* = -\frac{dy}{p(y)}.
\ee
Since $p(\xi_i) = 0$, the range $y_* \in (-\infty,\infty)$ covers only
the region between two horizons, and therefore a separate
$(t,y_*,x,\varphi)$ coordinate system is required for each `Rindler
diamond' in the spacetime (figure \ref{hor_diag}). Attention will be
focused on the first half of the scattering problem -- propagation
through a single Rindler diamond from $H_{al}^-$ and $H_{br}^-$ to
$H_{ar}^+$ and $H_{bl}^+$. The second half is subsequently found to be
trivial.

Using the form of the metric polynomial given in equation
(\ref{metpoly2}) the coordinate $y_*$ may be written as 
\ba
y_* & = & \frac{1}{\eta \zeta} \int \frac{dy}{ \{ y - \xi_1 \} \{ y -
\xi_2 \} \{ y - \xi_3 \} \{ y - \xi_4 \} } \nonumber \\[1.0mm]
 & = & -\gamma_1 \ln | \xi_1 - y | + \gamma_2 \ln | \xi_2 - y | -
\gamma_3 \ln | \xi_3 - y | + \gamma_4 \ln | \xi_4 - y |
\ea
where the $\gamma_i$ are constant positive functions of the four roots
$\xi_i$. In the `point-particle' limit, they become
\be
\gamma_i = \frac{1}{2} \pm \zeta + \mathcal{O}(\zeta^2),
\label{gammas}
\ee
with a `$-$' for $i = 1$ and $4$ and a `$+$' for $i = 2$ and
$3$. Clearly then, $y_* \rightarrow \infty$ at the acceleration and
inner black hole horizons, and $y_* \rightarrow -\infty$ at the outer
black hole horizon. 

Multiplying through by $p(y)$ and further recognising $\partial_{y_*}
= -p(y) \partial_y$ allows equation (\ref{Y_EQ}) to be re-written as
the one-dimensional wave equation 
\be
\left( \frac{\partial^2}{\partial y_*^2} + \left\{ \omega +
\frac{i}{2} s p'(y) \right\}^2 + V_{\mathrm{eff}} \right) \psi_s(y) = 0
\ee
where the effective potential is given by
\be
V_{\mathrm{eff}} = p(y) \left\{ \frac{1}{6} (2s^2+1) p''(y) + \Gamma_s
\right\}.
\ee
Near the outer black hole horizon $p(y) \rightarrow 0$ and hence the
effective potential becomes negligibly small. In addition $p'(y)
\rightarrow -2 \kappa$ where $\kappa$ is the surface gravity of the
horizon, and so the solution approaches
\be
\left. \psi_s(y) \right|_{y \rightarrow \xi_2} \sim \mathrm{e}^{
\pm i \{ \omega - i s \kappa \} y_*}.
\ee
Similarly, near the acceleration horizon $p'(y) \rightarrow 2 \kappa$,
and so the solution there approaches
\be 
\left. \psi_s(y) \right|_{y \rightarrow \xi_3} \sim \mathrm{e}^{
\pm i \{ \omega + i s \kappa \} y_*}.
\ee

It is clear that for some values of $s \neq 0$ the asymptotic limits
of these solutions are unbounded due to terms of the form
$\mathrm{e}^{\pm s \kappa y_*}$. Such singularities are of course not
physical, but instead are due to the fact that the terms $\psi_s(y)$
are Newman-Penrose quantities, and hence projections of some physical
quantity onto a tetrad which is ill-defined at the horizons. They are
therefore gauge- or frame-dependent. This effect may be
illustrated by re-expressing the tetrad in null coordinates and
subsequently applying a position dependent boost to demonstrate that
certain `well behaved' observers at the horizons will not witness such
pathology in all of the field quantities. 

In ingoing coordinates $(v,r,x,\varphi)$ defined via the relations 
\be
dv = \frac{1}{A} \{ dt + dy_* \}
\ee
for the null coordinate, and 
\be
dr = -\frac{1}{A \{x-y\}^2} \{ p(y)dy_* + dx \}
\ee
for the radial coordinate (\ref{r_coord}), the $l$ and $n$ tetrad
components may be written as   
\be
l^a = [ \{x-y\}\Lambda, \{x-y\}^{-1}\Lambda^{-1}, 0, 0 ] \qquad
\mbox{and} \qquad n^a = [ 0, -\{x-y\}^{-1}\Lambda^{-1}, 0, 0 ] 
\ee 
where
\be
\Lambda =  i \sqrt{2/p(y)}.
\ee
Both components remain singular at $p(y) = 0$, but by applying a
boost $\Lambda^{-1}$ they become~\cite{JS1}
\be
l^a = [ \{x-y\}, \{x-y\}^{-1}\Lambda^{-2}, 0, 0 ] \qquad \mbox{and}
\qquad n^a = [ 0, -\{x-y\}^{-1}, 0, 0 ]
\ee
which are evidently `well behaved' at the horizons. 

The effect of this boost is to transform the Newman-Penrose quantities
according to the relation 
\be
\psi_s^{\mathrm{in}}(y) = \Lambda^{-s} \psi_s(y),
\ee
for which the $y$ equation becomes
\ba
\lefteqn{\Bigg( {\partial \over \partial y} \left\{ p(y) {\partial
\over \partial y} \right\} - s p'(y){\partial \over \partial y} + {1
\over p(y)} \{ \omega^2 + i \omega s p'(y) \} } \nonumber \\[0.8mm]
& & \hspace{25mm} + {1 \over 6} ( 2s^2-3s+1)p''(y) + \Gamma_s \Bigg)
\psi_s^{\mathrm{in}}(y) = 0.
\ea
The asymptotic form of this equation may once again be obtained with
multiplication through by $p(y)$ to yield
\be
\left( {\partial^2 \over \partial y_*^2} + s p'(y) {\partial \over
\partial y_*} + \{\omega^2 + i \omega s p'(y) \} +
\hat{V}_{\mathrm{eff}} \right) \psi_s^{\mathrm{in}}(y) = 0,
\ee
the solutions to which are
\be
\left. \psi_s^{\mathrm{in}}(y) \right|_{y \rightarrow \xi_2} \sim
\mathrm{e}^{ -i \omega y_*} \mbox{\hspace{2mm} or \hspace{2.5mm}}
\mathrm{e}^{\{i \omega + 2 s \kappa \} y_*}
\ee
and 
\be
\left. \psi_s^{\mathrm{in}}(y) \right|_{y \rightarrow \xi_3} \sim
\mathrm{e}^{ -i \omega y_*} \mbox{\hspace{2mm} or \hspace{2.5mm}}
\mathrm{e}^{\{i \omega - 2 s \kappa \} y_*}.
\ee

In outgoing coordinates $(u, r, x, \varphi)$ where $u$ is defined by
\be
du = \frac{1}{A} \{ dt - dy_* \},
\ee 
the tetrad components $l$ and $n$ are rendered `well behaved' at the
horizons by applying a boost of $\Lambda$, the effect of which is to
transform the Newman-Penrose quantities according to 
\be
\psi_s^{\mathrm{out}}(y) = \Lambda^s \psi_s(y).
\ee
In this case the asymptotic form of the $y$ equation is given by 
\be
\left( {\partial^2 \over \partial y_*^2} - s p'(y) {\partial \over
\partial y_*} + \{\omega^2 + i \omega s p'(y) \} +
\tilde{V}_{\mathrm{eff}} \right) \psi_s^{\mathrm{out}}(y) = 0,
\ee
such that
\be
\left. \psi_s^{\mathrm{out}}(y) \right|_{y \rightarrow \xi_2} \sim
\mathrm{e}^{ i \omega y_*} \mbox{\hspace{2mm} or \hspace{2.5mm}}
\mathrm{e}^{\{-i \omega + 2 s \kappa \} y_*}
\ee
and 
\be
\left. \psi_s^{\mathrm{out}}(y) \right|_{y \rightarrow \xi_3} \sim
\mathrm{e}^{ i \omega y_*} \mbox{\hspace{2mm} or \hspace{2.5mm}}
\mathrm{e}^{\{-i \omega - 2 s \kappa \} y_*}.
\ee

The asymptotic behaviour of the field quantities expressed in both the
symmetric tetrad and the in- and outgoing tetrads clearly exhibit the
peeling behaviour expected from the peeling theorem of Newman \&
Penrose~\cite{NP1}. Ingoing observers `pick out' the ingoing component
of the field to be `non-special' -- neither singular nor identically
zero at the horizons~\cite{TEUK1}, and likewise the outgoing observers
`pick out' the outgoing components. The apparent pathology in some field
components is merely a result of some gauge choice, and
choosing to work in any particular gauge, with appropriately selected
boundary conditions, cannot therefore affect the final results. 

Care must nonetheless be exercised when attempting to define certain
quantities since $y$ equations such as (\ref{Y_EQ}) describe the
evolution of only a single spin-weight quantity $\psi_s(y)$. The
failure of these equations to be self-adjoint in cases other than $s =
0$ is manifested as an apparent lack of unitarity in the evolution of
any one spin-weight component. However, for fields with $s \neq 0$
there must also exist relations between the components of maximum
spin-weight -- the Teukolsky-Starobinsky identities~\cite{STARCH}, and
consequently the definitions of quantities such as conserved currents
must necessarily involve both spin-weight components of a
field~\cite{TEUK2}. 

Having laid the groundwork, the definition of asymptotic boundary
conditions for the desired scattering problem may be achieved by
analogy with a simple one-dimensional problem of scattering in a
potential well. Considering the initial hypersurfaces $H_{br}^-$ and
$H_{al}^-$, Hawking \& Ross argued~\cite{ROSS1} that the effects of
scattering a field from $H_{br}^-$ must be the image under left-right
interchange to the effects of scattering a field from
$H_{al}^-$. Since the metric must be symmetric under this exchange, 
calculating the scattering of a field incident only from the past
black hole horizon is justified since the reflection and transmission 
coefficients would be identical had the other choice been made
instead.  

With this in mind, an asymptotically-plane wave outgoing from
$H_{br}^-$ will propagate through the right Rindler diamond and
subsequently interact with some potential in the interior region. The
result of this interaction must then be an ingoing reflected component
through $H_{bl}^+$ and an outgoing transmitted component across
$H_{ar}^+$. These conditions may be written in symmetric tetrad
components as  
\be
\psi_s(y) = \mathrm{e}^{ \{ i \omega + s \kappa \} y_* } + C_R(s)
\mathrm{e}^{ \{ -i \omega - s \kappa \} y_* }
\label{BH_BOUND}
\ee
at the black hole horizon ($y_* \rightarrow - \infty$), and
\be
\psi_s(y) = C_T(s) \mathrm{e}^{ \{ i \omega - s \kappa \} y_* }
\label{ACC_BOUND}
\ee
at the acceleration horizon ($y_* \rightarrow \infty$), where $C_R(s)$
and $C_T(s)$ are related to the reflection and transmission
coefficients respectively, which remain to be determined. For the
degenerate case of $s=0$, the boundary conditions reduce to those
chosen for the massless scalar field in Hawking \& Ross, although the
`tortoise' coordinate is there defined with opposite sign.

\sect{Solving the \mbox{\boldmath $y$} Equation.} 

Having established physically consistent boundary conditions at the
horizons, the remaining problem of determining the reflection and
transmission coefficients reduces to that of solving
(\ref{Y_EQ}) subject to the constraints (\ref{BH_BOUND}) and
(\ref{ACC_BOUND}). Since this expression possesses the same regular
singular points encountered in the $x$ equation, attempting to find an
analytic solution will obviously necessitate recourse to the now
familiar `point-particle' limit.

Defining a new field $f_s(y)$ such that
\be
\psi_s(y) = \mathrm{e}^{\{i \omega - s \kappa \} y_*} f_s(y)
\ee
yields a slightly more pellucid form for this problem, in which
(\ref{Y_EQ}) may be re-written as
\ba
\lefteqn{\Bigg( \frac{\partial}{\partial y} \left\{ p(y) \frac
{\partial} {\partial y} \right\} - 2\left\{ i \omega - s \kappa
\right\} \frac{\partial}{\partial y} + \frac{1}{6} (2s^2+1)p''(y) +
\Gamma_s} \nonumber \\[0.8mm] 
& & \mbox{\hspace{10mm}} + \frac{1}{p(y)} \left\{ (i \omega - s \kappa
)^2 - (i \omega - \frac{1}{2} s p'(y))^2 \right\} \Bigg) f_s(y) = 0.
\label{NEW_Y_EQ}
\ea
The boundary conditions then become
\be
f_s(y) = \mathrm{e}^{ 2 s \kappa y_*} + C_R(s) \mathrm{e}^{-2 i \omega
y_*}
\label{bound_minus}
\ee
at the black hole horizon ($y_* \rightarrow -\infty$), and
\be
f_s(y) = C_T(s)
\label{bound_plus}
\ee
at the acceleration horizon ($y_* \rightarrow \infty$).

In moving to the `point-particle' limit this equation is further
simplified by changing coordinates
\be
y \rightarrow \hat{y} = \frac{1}{\beta} \left( y - \frac{1}{2} \{
\xi_3 + \xi_2 \} \right)
\ee
where
\be
\beta \stackrel{\mathrm{def}}{=} \frac{1}{2} \{ \xi_3 - \xi_2 \} =
\frac{1}{2 \zeta} \{ 1 + \zeta + \mathcal{O}(\zeta^2) \}.
\ee
The range of $y$ is then mapped to $\hat{y} \in [-1,1]$, and the
symmetry imposed on the metric polynomial allows it to be re-written
as
\be
p(\hat{y}) = -\eta \zeta \beta^4 \{ \hat{y}^2 - \alpha^2 \} \{
\hat{y}^2 - 1 \}
\ee
with
\be
\alpha \stackrel{\mathrm{def}}{=} \frac{ \xi_4 - \xi_1 }{ \xi_3 -
\xi_2 } = 1 + 4 \zeta + \mathcal{O}(\zeta^2).
\ee

With this change in place the three polynomial functions $p(y)$,
$p'(y)$ and $p''(y)$ -- where $'$ denotes differentiation with respect
to $y$ -- may be expressed as power series in $\zeta$ which are
respectively 
\ba
p(\hat{y}) & = & -\frac{1}{16 \zeta^2} (1-\hat{y}^2) \bigg\{ (1 -
\hat{y}^2) + 2(1 + 3\hat{y}^2) \zeta + \mathcal{O}(\zeta^2) \bigg\},
\\[0.8mm]
p'(\hat{y}) & = & \frac{1}{2 \zeta} \hat{y} \bigg\{ (1-\hat{y}^2) - (1
- 5\hat{y}^2) \zeta + \mathcal{O}(\zeta^2) \bigg\}, \\[0.8mm]
p''(\hat{y}) & = & \bigg\{ (1-3\hat{y}^2) + 12\hat{y}^2 \zeta +
\mathcal{O}(\zeta^2) \bigg\}.
\ea
If $(\hat{y}^2-1) \sim \mathcal{O}(1)$ then only the first terms in
these series need be retained. Substituting them into (\ref{NEW_Y_EQ})
with the identification (\ref{gamma_s}) for $\Gamma_s$, and further
assuming $\omega \sim \mathcal{O}(1)$ gives
\be
\left( \frac{ \partial }{ \partial \hat{y} } \left\{ (\hat{y}^2 - 1)^2
\frac{ \partial }{ \partial \hat{y} } \right\} + 2 \left\{ \hat{y}^2 -
[1 + 2j(j+1)] \right\} \right) f_s(\hat{y}) = 0
\label{mid_range}
\ee
where $\partial_{\hat{y}} = \beta \partial_y$. In this approximate
equation $j$ is used in favour of $j_0$ since regarding $j$ as a
half-integer would introduce degeneracies that are not present in the
exact equation. The solution may be expressed in the form
\be
f_s(\hat{y}) = \frac{1}{ \{ \hat{y}^2-1\}^{j+1}} \left\{ 
A \{ \hat{y}+1 \}^{2j+1} + B \{ \hat{y}-1 \}^{2j+1} \right\}
\label{centre_sol}
\ee
for arbitrary constants $A$ and $B$. Clearly this function diverges at
the boundaries $\hat{y} = \pm 1$, but this is of little real concern
since (\ref{mid_range}) is not valid near those points. A separate
approximation must therefore be used in the neighbourhoods of the
boundaries, which may be obtained by defining a new coordinate $q$
where 
\be
\hat{y} = \pm \{ 1 - 4 \zeta q \}.
\ee

Dealing first with the neighbourhood of $\hat{y} = -1$, the leading
order terms in equation (\ref{NEW_Y_EQ}) become
\be
\left( \frac{\partial}{\partial q} \left\{ q(q+1)
\frac{\partial}{\partial q} \right\} + \{ i \omega - s \}
\frac{\partial}{\partial q} - \frac{i \omega s}{q} - j(j+1)
\right) f_s(q) = 0.
\ee
Searching for a series solution of the form $f_s(q) = \sum
a_{\lambda}q^{k + \lambda}$ yields indicial roots $k = +s$ and $-i
\omega$, and a general one-term recurrence relation of the form
\be
a_{n+1} = a_n \left\{ \frac{ j(j+1) - (k+n)(k+n+1)}{ (k+n+1)[k+n+i
\omega -s+1] - i \omega s } \right\}.
\ee
This may be identified as a ${ {}_2 F_1 }$ Gaussian hypergeometric
series, convergent throughout the unit disk $|q|<1$ for the
non-integer parameters considered here. The solution obeying the
boundary condition (\ref{bound_minus}) is therefore
\ba
\lefteqn{ f_s(q) = \{4 \zeta q\}^{s} F \Big( \{j+s+1\},\{-j+s\};\{i
\omega +s+1\}; -q \Big) } \nonumber \\[0.8mm]
& & \mbox{\hspace{15mm}} + C_R(s) \{4 \zeta q\}^{-i \omega} F \Big( \{j -
i \omega +1\},\{-j - i \omega \};\{-s - i \omega+1\}; -q \Big).
\label{bh_sol}
\ea

Moving to the neighbourhood of $\hat{y} = +1$, the leading-order terms
in equation (\ref{NEW_Y_EQ}) now give
\be
\left( \frac{\partial}{\partial q} \left\{ q(q+1)
\frac{\partial}{\partial q} \right\} - \{ i \omega - s \}
\frac{\partial}{\partial q} + \frac{i \omega s}{q+1} - j(j+1)
\right) f_s(q) = 0.
\label{plus_one}
\ee
Written in this form an attempted series solution generates a
two-term recurrence relation which is of little use. However, a one-term
relation may be generated by instead considering the coordinate
$\tilde{q} \stackrel{\mathrm{def}}{=} q+1$. In this case
\be
\left( \frac{\partial}{\partial \tilde{q}} \left\{ \tilde{q}(\tilde{q}
- 1) \frac{\partial}{\partial \tilde{q}} \right\} - \{ i \omega - s \}
\frac{\partial}{\partial \tilde{q}} + \frac{i \omega s}{\tilde{q}} -
j(j+1) \right) f_s(\tilde{q}-1) = 0
\ee 
which again gives indicial roots of $k = +s$ and $-i \omega$, but with
a general recurrence relation of the form
\be
a_{n+1} = a_n \left\{ \frac{ (k+n)(k+n+1) - j(j+1) }{ (k+n+1)[k+n + i
\omega -s+1] - i \omega s } \right\}.
\ee
Naturally this may also be identified as a ${ {}_2 F_1 }$ Gaussian
hypergeometric series, and the general solution for $f_s(q)$ may be
expressed as
\ba
\lefteqn{ \{q+1\}^{-s} f_s(q)  = C F \Big( \{j+s+1\}, \{-j+s\}; \{i
\omega +s+1\}; q+1 \Big)} \nonumber \\[0.8mm]
& & \mbox{\hspace{15mm}} + D \{q+1\}^{-s-i \omega} F \Big( \{j-i \omega
+1\}, \{-j-i \omega\}; \{-s-i \omega +1\}; q+1 \Big)
\label{tilde_eqn}
\ea
for arbitrary constants $C$ and $D$. If these are chosen
respectively as
\ba
C & = & C_T(s) \times \frac{ \Gamma(s-i\omega+1) \Gamma(-s-i\omega) } {
\Gamma (-j-i\omega) \Gamma(j-i\omega+1) } \\[0.8mm]
D & = & C_T(s) \times \frac{ \Gamma(s-i\omega+1) \Gamma(s+i\omega) } { 
\Gamma (-j+s) \Gamma(j+s+1) },
\ea
then (\ref{tilde_eqn}) may be transformed (see for
example~\cite{table}) to a solution of the form
\be
f_s(q) = C_T(s) \{q+1\}^{s} F \Big( \{j+s+1\}, \{-j+s\}; \{s-i \omega
+1\}; -q \Big)
\label{acc_sol}
\ee
which clearly satisfies the boundary condition (\ref{bound_plus}).

At this stage it should be noted that equation (\ref{plus_one}) and
its solution (\ref{acc_sol}) will also be valid for the entire region
$\hat{y} \in [1,\alpha]$. To zeroth order this solution therefore
describes the form of the incident field with the desired boundary
condition on $H_{ar}^+$, restricted to zero on $H_{al}^+$ , through
the top Rindler diamond bounded by $\mathcal{J}^+$. $C_T(s)$ is
therefore related to the transmission coefficient for the complete
scattering problem from the previously defined initial Cauchy surface
to the future Cauchy surface.

\sect{Transmission and Reflection Coefficients}

Although the definition of the Gaussian hypergeometric series ensures
its convergence only within the unit disk, there nonetheless exist
transformation formulae for analytically continuing to the region $|q|
>1$~\cite{table}. The boundary neighbourhood solutions (\ref{bh_sol})
and (\ref{acc_sol}) may therefore be analytically continued to large
values of $q$ and subsequently matched to the function
(\ref{centre_sol}) which is valid in this `central' region.

In the interest of notational convenience, the following factors are
defined: 
\ba
U & = & \displaystyle{ \Gamma(-s-i\omega+1)
\Gamma(2j+1) \over \Gamma(j-i\omega+1) \Gamma(j-s+1) } , \\[0.8mm]
V & = & \displaystyle{ \Gamma(s+i\omega+1)
\Gamma(j-i\omega+1) \Gamma(j-s+1) \over \Gamma(j+s+1) \Gamma(-s-i\omega+1)
\Gamma(j+i\omega+1) }, \\[0.8mm]
W & = & \displaystyle{ \Gamma(s-i\omega+1)
\Gamma(-2j-1) \over \Gamma(-j-i\omega) \Gamma(-j+s) }, \\[0.8mm]
X & = & \displaystyle{ \Gamma(-s-i\omega+1)
\Gamma(-2j-1) \over \Gamma(-j-i\omega) \Gamma(-j-s) }, \\[0.8mm]
Y & = & \displaystyle{ \Gamma(s+i\omega+1)
\Gamma(-j-i\omega) \Gamma(-j-s) \over \Gamma(-j+s) \Gamma(-s-i\omega+1)
\Gamma(-j+i\omega) }, \\[0.8mm]
Z & = & \displaystyle{ \Gamma(s-i\omega+1)
\Gamma(2j+1) \over \Gamma(j-i\omega+1) \Gamma(j+s+1) },
\ea
where it is clear from their definitions that, amongst others, the
following relations hold:
\be
\begin{array}{cccccc}
U(-s) & = & Z(s), \mbox{\hspace{10mm}} & W(-s) & = & X(s), \nonumber \\[0.8mm]
V^*(-s) & = & V^{-1}(s), \mbox{\hspace{10mm}} & Y^*(-s) & = & Y^{-1}(s). \nonumber
\label{ptm1}
\end{array}
\ee
With the benefit of these definitions, the analytic continuation of
$f_s(q)$ from the black hole horizon may be written as
\ba
\lefteqn{ f_s(q) =  X \left( \{4 \zeta\}^{-i\omega} C_R(s) + \{4 \zeta\}^{s} Y
\right) \{q\}^{-j-1} } \nonumber \\[0.8mm]
& & \mbox{\hspace{20mm}}  + U \left( \{4 \zeta\}^{-i\omega} C_R(s) +
\{4 \zeta\}^{s} V \right) \{q\}^j
\label{cont_bh}
\ea
for $q \gg 1$. Likewise the analytic continuation of $f_s(q)$ from the
acceleration horizon becomes
\be
f_s(q) = W C_T(s) \{q\}^{-j-1} + Z C_T(s) \{q\}^j. 
\label{cont_acc}
\ee

By substituting for the coordinate $q$ defined near $\hat{y}
= \pm 1$, equation (\ref{centre_sol}) becomes
\ba
\lefteqn{ f_s(q) \approx \Big( \{-1\}^{-j-1} \times 2^{j-1} \times A \times \{
\zeta q \}^j \Big) } \nonumber \\[0.8mm]
& & \mbox{\hspace{20mm}} + \Big( \{-1\}^j \times 2^{-j-2} \times B
\times \{ \zeta q \}^{-j-1} \Big) 
\ea
by the black hole horizon, and
\ba
\lefteqn{ f_s(q) \approx \Big( \{-1\}^{-j-1} \times 2^{-j-2} \times A \times \{
\zeta q \}^{-j-1} \Big) } \nonumber \\[0.8mm]
& & \mbox{\hspace{25mm}} + \Big( \{-1\}^j \times 2^{j-1} \times B
\times \{ \zeta q \}^j \Big)
\ea
by the acceleration horizon. Matching these results with the
corresponding analytically continued expressions (\ref{cont_bh}) and
(\ref{cont_acc}) requires for consistency the relations
\be
U \left( C_R(s) + \{4\zeta\}^{s} V \right) = \{2 \zeta\}^{2j+1} W C_T(s)
\ee
and
\be
X \left( C_R(s) + \{4\zeta\}^{s} Y \right) = \left\{ \frac{1}{2 \zeta}
\right\}^{2j+1} Z C_T(s)  
\ee
where the phase factor $\{4 \zeta\}^{-i\omega}$ has been absorbed into
$C_R(s)$. These may obviously be solved for the two unknown
factors. By further defining
\be 
\alpha = \{ 2 \zeta \}^{2 j +1}
\ee
and
\be
\beta = \frac{UZ}{WX}
\ee
they may be written as
\be
C_R(s) = \frac{\alpha^2\beta Y - V}{1 - \alpha^2 \beta} \times \{4
\zeta \}^{s}
\ee
and
\be
C_T(s) = \alpha \frac{X}{Z} \frac{Y - V}{1 - \alpha^2 \beta} \times
\{4 \zeta \}^{s}.
\ee  

Appropriate use of the Teukolsky-Starobinsky identities for the
components of extreme spin-weight allows the verification of current
conservation, but this relative normalisation technique may be
bypassed by instead calculating $C_R(-s)$ and
$C_T(-s)$~\cite{PAGE}. With application of the relations (\ref{ptm1})
between the various factors under a replacement of $s \rightarrow -s$,
it is clear that, as required for consistency,  
\be
\Gamma_R + \Gamma_T = 1
\ee
if 
\be
\Gamma_R \stackrel{\mathrm{def}}{=} | C_R(s) C_R(-s) | 
\ee
and
\be
\Gamma_T \stackrel{\mathrm{def}}{=} \pm | C_T(s) C_T(-s) | 
\ee
where the `$+$' is necessary for bosonic fields, and the `$-$' for
fermionic fields. To leading order in $\zeta$ this definition of the
transmission coefficient reduces to
\be
\Gamma_T = \pm \alpha^2 | \beta | \left\{ 2 - Y/V - V/Y \right\} +
\mathcal{O}(\alpha^4).
\ee
Since $s$ is a discrete parameter, the term $Y/V$ may be written as 
\be
Y/V = \pm \frac{\sin \{\pi(j-i\omega)\}}{\sin
\{\pi(j+i\omega)\}} \stackrel{\mathrm{def}}{=} \pm \exp ( i \theta )
\ee
where again the `$+$' is required for bosonic fields and the `$-$' for
fermionic fields. Exploiting this definition $\Gamma_T$ becomes, after
a little algebra
\be
\Gamma_T = -2 \alpha^2 \left\{ \frac{ \Gamma(-2j-1) \Gamma(j+s+1) }{
\Gamma(-j+s) \Gamma(2j+1) } \right\}^2 \left| \frac{ \Gamma(j+i \omega
+1)}{\Gamma(-j+ i \omega)} \right|^2 (-1 \pm \cos \theta ) +
\mathcal{O}(\alpha^4). 
\ee

By assumption $j = j_0 + \mathcal{O}(\zeta)$ where $j_0$ is a
non-negative half-integer, and hence $j_0 = l_0 + |s|$ for
integer $l_0 \ge 0$. Thus, in a leading order calculation, the limit
may be taken in which $j-s$ tends to a non-negative integer. The first
term may be re-expressed as
\be
\left\{ \frac{ \Gamma(-2j-1) \Gamma(j+s+1)} { \Gamma(-j+s)
\Gamma(2j+1)} \right\}^2 = \left\{ \frac{ \Gamma(j+s+1) \Gamma(j-s+1)}
{ \Gamma(2j+1) \Gamma(2j+2)} \right\}^2 \times \left[ \frac{ \sin
\{\pi (j-s)\}}{\sin \{\pi (2j+1)\}} \right]^2
\ee
which in this limit becomes
\be
\frac{1}{4} \left\{ \frac{ (j+s)! (j-s)!}{(2j)! (2j+1)!} \right\}^2
\ee
for all spins. The second term may be written as a finite product by
recognising that
\be
\left| \frac{\Gamma(n+i \omega+1)}{\Gamma(-n+i \omega)} \right|^2 =
\left\{ 
\begin{array}{ll}
{\displaystyle \frac{1}{\omega^2} \prod_{i=0}^{n} (i^2+n^2)^2} &
\mbox{n integer,} \\ 
{\displaystyle \prod_{i = 1/2}^{n} (i^2+n^2)^2} & \mbox{n half-integer.}
\end{array}
\right.
\ee

Finally, by expanding the definition of $\theta$ it is clear that
$\cos \theta \rightarrow -1$ for bosonic fields while $\cos \theta
\rightarrow +1$ for fermionic fields, and so the transmission
coefficient becomes
\be
\Gamma_T = \{2 \zeta \}^{4j+2} \left\{ \frac{ (j+s)! (j-s)!}{(2j)!
(2j+1)!} \right\}^2 \frac{1}{\omega^2} \prod_{i=0}^{j_0}
(i^2+\omega^2)^2 + \cdots
\ee
for integer $s$, and
\be
\Gamma_T = \{2 \zeta \}^{4j+2} \left\{ \frac{ (j+s)! (j-s)!}{(2j)!
(2j+1)!} \right\}^2 \prod_{i=1/2}^{j_0} (i^2+\omega^2)^2 + \cdots
\ee
for half-integer $s$, where `$\cdots$' denotes terms of higher order
in $\zeta$.

Clearly the largest contribution will come from the $j_0 = |s|$ ($l_0
= 0$) modes, for which the expressions simplify to
\be
\Gamma_T = \left\{
\begin{array}{ll}
{\displaystyle \{2 \zeta\}^2 \omega^2} & \mbox{spin 0,} \\[1.0mm]
{\displaystyle \frac{1}{4}\{2 \zeta\}^4 ( 1/4 + \omega^2)^2} &
\mbox{spin 1/2,} \\[2.5mm]
{\displaystyle \frac{1}{36} \{2 \zeta \}^6 \omega^2 ( 1+\omega^2)^2} &
\mbox{spin 1.}
\end{array} \right.
\ee
This leading order approximation, valid for $\omega \leq 1$, may also
be extended to spin $2$ fields incident on the vacuum C metric,
although it would appear plausible to expect the overall power of
$\zeta$ in this approximation to remain $4s+2$ when in addition these
fields are considered incident on the electrovac solution. 

\sect{Discussion}
In calculating an estimate of the affect of Planck size virtual black
hole loops on coherence loss in incident scalar fields, Hawking \&
Ross~\cite{ROSS1} exploited the separability of the conformally
invariant scalar wave equation in one particular metric -- the
electrovac C metric. The scattering calculation is performed in the
Lorentzian section , by considering the propagation of a field from an
initial Cauchy surface to a future Cauchy surface, suitably
constructed from the left and right black hole and acceleration
horizons and $\mathcal{J}^+$. The total scattering problem is split
into two distinct sections: propagation through the right Rindler
diamond to define reflection and transmission coefficients, followed
by propagation through the top Rindler diamond to $\mathcal{J}^+$
which is found to be trivial. The transmission factor $C_T$ (which,
modulus an arbitrary phase factor, may be identified with
$\sqrt{\Gamma_T}|_{s=0}$) is necessary in calculating the Bogoliobov
coefficients for this scattering problem and hence in estimating the
number density $N_{\omega}$ of particles at $\mathcal{J}^+$.

This procedure has herein been repeated and extended to fields of spin
$\frac{1}{2}$ and $1$ by finding and separating the Teukolsky
equations on the electrovac C metric. (As a byproduct the resulting
equations are in addition valid for fields of spin $0$, $\frac{1}{2}$,
$1$, and $2$ incident on the vacuum C metric). It is shown that the
angular parts of these solutions comprise a complete set of orthogonal
polynomials ${}_s\phi_{jm}(x) \mathrm{e}^{i m \varphi}$ characterised
by two real numbers $j$ and $m$ and a half-integer spin-weight
parameter $s$. In the `point-particle' limit these functions reduce to
the spin-weighted spherical harmonics ${}_s Y_{j_0 m_0} (\theta,
\varphi)$ with $j_0$ and $m_0$ half-integer. 

A suitable generalisation of the radial boundary conditions considered
in Hawking \& Ross is subsequently demonstrated to be appropriate
despite the appearance of apparent pathologies in the asymptotic
behaviour of non-zero spin field quantities. These effects are shown
to be gauge sensitive rather than physical in nature. Fixing a gauge
(or frame) necessitates careful handling of frame independent
quantities such as conserved currents, which na\"{\i}vely appear not to
be conserved since the radial equation is not self-adjoint.

In the chosen gauge the radial solutions approach ${}_2 F_1$
hypergeometric functions close to the horizons, and these are matched
to the solution in the interior region to fix the transmission and
reflection factors $C_T(s)$ and $C_R(s)$. The hypergeometric
approximation is in addition shown to be valid throughout the top
Rindler diamond and so the two factors $C_T(s)$ and $C_R(s)$
determine the complete scattering problem from the initial Cauchy
surface to the future Cauchy surface.

By suitably defining the transmission and reflection coefficients such
that they are invariant under the transformation $s \rightarrow -s$,
current conservation may be demonstrated, in that $\Gamma_R + \Gamma_T
= 1.$ The leading order contribution to the transmission coefficient
is subsequently shown to behave as $\zeta^{4s+2}$ for $\omega \leq 1$,
where $\zeta$ is a small parameter. It is clear therefore that 
transmission is suppressed for fields of higher spin.

Calculations for an incident scalar field indicated that a finite
non-zero number of particles should be detected at $\mathcal{J}^+$,
implying loss of quantum coherence since each of these particles may
be thought of as one of a virtual pair whose partner has fallen into
the black hole. Since transmission is suppressed for higher spin fields
this effect will be correspondingly reduced. A full derivation of this
result will be given in a subsequent paper.

\end{document}